\documentclass[11pt]{article}
\usepackage{graphicx}

\begin{document}

\def\xslash#1{{\rlap{$#1$}/}}
\def \p {\partial}
\def \dd {\psi_{u\bar dg}}
\def \ddp {\psi_{u\bar dgg}}
\def \pq {\psi_{u\bar d\bar uu}}
\def \jpsi {J/\psi}
\def \psip {\psi^\prime}
\def \to {\rightarrow}
\def\bfsig{\mbox{\boldmath$\sigma$}}
\def\DT{\mbox{\boldmath$\Delta_T $}}
\def\xit{\mbox{\boldmath$\xi_\perp $}}
\def \jpsi {J/\psi}
\def\bfej{\mbox{\boldmath$\varepsilon$}}
\def \t {\tilde}
\def\epn {\varepsilon}
\def \up {\uparrow}
\def \dn {\downarrow}
\def \da {\dagger}
\def \pn3 {\phi_{u\bar d g}}

\def \p4n {\phi_{u\bar d gg}}

\newcommand{\pslash}{p \hspace{-1.0ex}\slash}

\def \bx {\bar x}
\def \by {\bar y}

\begin{center}
{\Large\bf A Note on the $k_T$-Factorization for Exclusive Processes}
\par\vskip20pt
F. Feng$^{1}$, J.P. Ma$^{2,3}$  and Q. Wang$^{4}$    \\
{\small {\it $^1$ Theoretical Physics Center for Science Facilities, Institute of High Energy Physics,
\\
Academia Sinica, Beijing 100049, China }} \\
{\small {\it $^2$ Institute of Theoretical Physics, Academia Sinica,
P.O. Box 2735,
Beijing 100190, China\\
$^3$ Center for High-Energy Physics, Peking University, 100080, China  \\
$^4$ Department of Physics and Institute of Theoretical
Physics, Nanjing Normal University, Nanjing, Jiangsu 210097, P.R.China }}
\end{center}
\vskip 1cm
\begin{abstract}
We show in detail that the $k_T$-factorization for exclusive processes is gauge-dependent and
inconsistent.
\vskip 5mm \noindent
\end{abstract}
\vskip 1cm
\par
The $k_T$-factorization has been widely used to study exclusive B-decays(see references in \cite{NLi,FMW0}).
In this factorization, transverse momenta $k_T$ of partons are taken into account.
The hard part of the factorization is extracted from scattering of off-shell partons and it depends
on the transverse momenta. Because the scattering is of off-shell partons, it is likely
that the hard part, hence the factorization,  is gauge-dependent. To our knowledge, the $k_T$-factorization
for exclusive B-decays has been not studied at one-loop completely.
\par
The first one-loop study of the $k_T$-factorization is for the case $\gamma^* + \pi \to \gamma$ in \cite{NLi},
where the hard part is calculated in Feynman gauge. It has been claimed
that the $k_T$-factorization is gauge-independent\cite{NLi}.
In \cite{FMW0} it has been pointed out that the $k_T$-factorization is in fact gauge-dependent because of
singular contributions from the wave functions to the hard part in a general covariant gauge.
A method has been suggested in \cite{LiM} to eliminate such singular contributions. However, such a
method is inconsistent as pointed out in \cite{FMW1}.
To determine the hard part at one-loop, one needs to calculate at one loop the form factor in the case and
the wave function. It should be noted that there are no complete
one-loop results of wave functions
in the general covariant gauge. Therefore, the gauge invariance of the $k_T$-factorization
has been never checked explicitly at one-loop with this gauge,
except the singular contribution found in \cite{FMW0} .

\par
Recently, the case $\gamma^* + \pi \to \pi$ has been studied with the $k_T$-factorization at one-loop in \cite{LiS},
where it has been pointed out that the gauge invariance of the $k_T$-factorization is proven in \cite{LiM}.
In this note we will study this issue at tree-level and beyond tree-level. We show  that the $k_T$-factorization is gauge-variant
and hence inconsistent.

\par\vskip10pt

{\bf 1.} Because a gluon is exchanged at tree-level, the problem of the gauge-invariance already
appears in the $k_T$-factorization for the form factor of the process $\gamma^* + \pi \to \pi$.
We use the  light-cone coordinate system, in which a
vector $a^\mu$ is expressed as $a^\mu = (a^+, a^-, \vec a_\perp) =
((a^0+a^3)/\sqrt{2}, (a^0-a^3)/\sqrt{2}, a^1, a^2)$ and $a_\perp^2
=(a^1)^2+(a^2)^2$.
We take a frame
in which the initial- and final $\pi$ have the momentum $P^\mu \approx (P^+, 0,0,0)$
and $K^\mu \approx (0,K^-,0,0)$, respectively. $P^+$ and $K^-$ are large and the square of the momentum transfer
is given by $Q^2=2 P^+ K^-$.
In the $k_T$-factorization
the form factor for  large $Q^2$ is written as:
\begin{equation}
F_\pi (Q) \sim \alpha_s \int \left [ dx d^2 k_{\perp}\right ] \left [ d y d^2  p_{\perp}\right]
 \phi (x, k_\perp) \phi (y, p_{\perp})
H(x,k_{\perp},y, p_\perp).
\end{equation}
In the above $\phi (x, k_\perp)$ and $\phi (y, p_\perp)$ are  the wave function for the initial-
and final $\pi$, respectively. These two wave functions can be different and their definitions
will be given later.
The hard part $H$ is determined by replacing the initial- and final $\pi$ with a off-shell quark pair respectively.
We take the quark pair $q(k_1) \bar q(P-k_1)$ for the initial $\pi(P)$ and
the pair $q(k_2) \bar q(K-k_2)$ for the final $\pi(K)$. The momentum $k_1$ and $k_2$ are specified as:
\begin{equation}
k^\mu_1 = (x_0 P^+,0, \vec k_{1\perp}), \ \ \ k_2^\mu = (0, y_0 K^-, \vec k_{2\perp}). \ \ \
\label{mom}
\end{equation}
We have here $k_1^2= -k^2_{1\perp} \neq 0$ and $k_2^2=-k^2_{2\perp}\neq 0$, reflecting the fact that the quark pairs are off-shell.
To determine $H$ one uses the quark pairs to calculate the form factor and wave functions.
For the off-shell quark pairs in the initial- and final state one uses
the spin projection $\gamma_5 \gamma^-$ and $\gamma^+ \gamma_5$ respectively.
With these projections one picks up the leading-twist contributions.
At tree-level the wave functions, denoted as $\phi^{(0)}$, are proportional
to $\delta$-functions and gauge-invariant because no gluon is exchanged.
It is straightforward to obtain in Feynman gauge $H$ at tree-level, denoted as $H^{(0)}$ as:
\begin{equation}
 H^{(0)} (x,k_\perp, y, p_\perp)  = \frac{1}{ x y Q^2 +(\vec p_\perp -\vec k_\perp )^2} .
\end{equation}
In deriving this result one has used the power counting: $x Q \sim  y Q \sim p_\perp \sim k_\perp\sim \delta$
and only the leading terms in $\delta$ has been taken into account.
\par
Because it is derived
in Feynman gauge with off-shell partons, $H$ can be different in different gauges.
Supposing we work in the general covariant gauge, in which the gluon propagator reads:
\begin{equation}
\frac{-i}{q^2+i\varepsilon} \left [ g^{\mu\nu} - \alpha \frac{ q^\mu q^\nu}{q^2+i\varepsilon} \right ]
\end{equation}
with the gauge parameter $\alpha$, we obtain $H$ as:
\begin{equation}
H^{(0)}(x,k_\perp,y, p_\perp) = \frac{1}{ x  y Q^2 +(\vec p_\perp
-\vec k_\perp )^2} \left ( 1  - \frac{\alpha}{2}\frac{\vec k_\perp \cdot
(\vec k_\perp -\vec p_\perp)} { x y Q^2 +(\vec p_\perp -\vec k_\perp )^2}
\right ) .
\end{equation}
It is clear that $H$ is gauge-dependent at tree-level.
The terms with $\alpha$ are at the same order of $H^{(0)}$ determined
in Feynman gauge. They are not suppressed by power of $\delta$.
Therefore, the gauge-dependent term
can not be neglected. Similar results for exclusive $B$-meson decays are also obtained\cite{WZT}.
Because the transverse momenta appear in numerators of the gauge-dependent terms, one may argue that these terms
may be factorized with higher-twist
operators other than the leading-twist operator used to defined $\phi$'s. If one can do so,
these terms are still gauge-dependent and can not be neglected with the power counting,
 in comparison with the
term factorized with $\phi$'s. This can be illustrated with the term in the numerator which is linearly in
$k_\perp$ in Eq.(5). The contribution can be factorized with a wave function
defined with the matrix element of the initial $\pi$ $\langle 0 \vert q(0) \gamma^5 \gamma^+ \partial^\mu_\perp
q(y) \vert \pi (P)\rangle$ with $y^+ =0$. In this case one may need to analyze
the contribution with the incoming off-shell quark pair combined a off-shell gluon. Adding this
it may result in that the derivative $\partial^\mu_\perp$ becomes the covariant derivative $D^\mu_\perp$.
But it is here not important how this term is factorized. The important is that the form factor
calculated with off-shell partons has a gauge-dependent and nonzero contribution.
In Eq.(5) we have factorized it
with the wave function. Regardless how it is factorized, this gauge-dependent and nonzero contribution
can not be eliminated by factorization with different operators.
Therefore, the $k_T$-factorization at tree-level is already
inconsistent.
\par
An interesting fact should be noted when one studies the $k_T$-factorization beyond tree-level
in Feynman gauge. In this gauge, the form factor and the wave functions
will not have  I.R. or collinear divergences, because they
are regularized by the off-shellness of partons, i.e.,  by $\ln k_{1\perp}^2$ and $\ln k_{2\perp}^2$ here.
 Since everything is finite, the factorization is
a trivial task.

\par\vskip20pt
\par
\begin{figure}[hbt]
\begin{center}
\includegraphics[width=10cm]{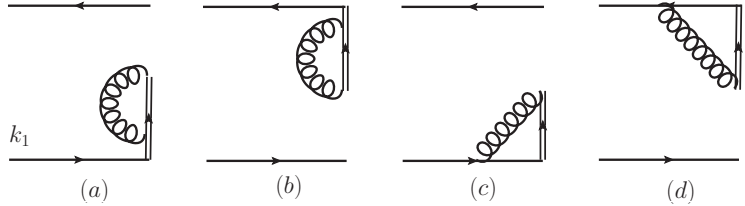}
\end{center}
\caption{The one-loop diagrams of the wave function.The double line represents the gauge link in Eq.(\ref{def}). }
\label{Feynman-dg1}
\end{figure}
\par
{\bf 2.} To determine the hard part at one-loop, one needs to calculate the form factor
and wave functions with the quark pairs at one-loop. With these one-loop results and by expanding
the right-hand side of Eq.(1) in $\alpha_s$ one can obtain $H^{(1)}$. $H^{(1)}$
will receive a contribution proportional to
\begin{equation}
  H^{(0)} \otimes \phi^{(1)} = \int_0^1 dx \int d^2 k_\perp H^{(0)}(x,k_\perp, y_0,k_{2\perp})
   \phi^{(1)}(x,k_\perp),
\end{equation}
where $\phi^{(1)}$ is the one-loop contribution to $\phi$ calculated with the off-shell quark pair.
The wave function $\phi$ of the incoming $\pi (P)$ is defined as:
\begin{eqnarray}
\phi(x, k_T,\zeta, \mu) = \int \frac{ d z^-d^2 z_\perp}{(2\pi )^3}
e^{ix P^+ z^- - i \vec z_\perp\cdot \vec
k_T}
\langle 0 \vert \bar q(0) L_u^\dagger (\infty, 0)
  \gamma^+ \gamma_5 L_u (\infty,z) q(z) \vert \pi(P) \rangle\biggr\vert_{z^+=0},
\label{def}
\end{eqnarray}
where $L_u(\infty,z)$ is a gauge link starting from the space-time point $z$
to $\infty$ along the vector $u$. The vector $u$ is defined as $u^\mu =(u^+,u^-,0,0)$.
In the above definition the limit $u^+ \to 0$ should
be taken, i.e., any term proportional $u^+$ should be neglected.
The limit is taken after all loop-integrations.
Besides $x$, $k_T$ and the renormalization
scale $\mu$ the wave function depends on the vector $u$ through the parameter
$\zeta^2 = 2 u^-(P^+)^2/u^+ \approx (2u\cdot P)^2/u^2$.
This results in that the hard part $H$ will also depend on $\zeta$. This $\zeta$-dependence
is very useful for resummation of double log's.
\par
Similarly, one can defined the wave function of the outgoing $\pi (K)$ with the gauge link
along the direction $u^{\prime \mu}=(u^{\prime +}, u^{\prime -}, 0,0)$ in the limit $u^{\prime -}\to 0$.
The wave function of $\pi(K)$ and hence the hard part will also depend on $\zeta'$, defined
as $\zeta^{\prime 2}= 2 u^{\prime +}(K^-)^2/u^{\prime -}  \approx (2 u^{\prime} \cdot K)^2/u^{\prime 2}$.
Therefore, the two wave functions are in general different. In practice one may take the choice
$\zeta=\zeta^{\prime}$. Our discussion in the following will not be affected if one makes or does not
make the choice.
\par
The wave function has been studied with an on-shell quark pair at one-loop in \cite{TMD1}. The obtained results
are gauge-invariant. But, the wave function in the $k_T$-factorization is calculated with an off-shell quark pair,
it is not gauge-invariant.
Hence, it is possible that the $\zeta$-dependence is also gauge-dependent. This in turn
gives to the hard part $H$ a gauge-dependent $\zeta$-dependence, since the form factor does not depend
on $\zeta$. If this is the case, it clearly indicates the gauge-variance of the $k_T$-factorization.
In the general covariant gauge, because the gauge-dependent term in Eq.(4) is proportional
to $q^\mu q^\nu$, it is very simple to show that this term does not give at one-loop the $\zeta$-dependence
extra than that in Feynman gauge. However, it is unclear if there exists a gauge-dependent $\zeta$-dependence beyond
one-loop. The situation changes if we work with an axial gauge, there is an extra $\zeta$-dependence
at one-loop. It is even worse that the extra $\zeta$-dependence is I.R. divergent.

\par
\begin{figure}[hbt]
\begin{center}
\includegraphics[width=8cm]{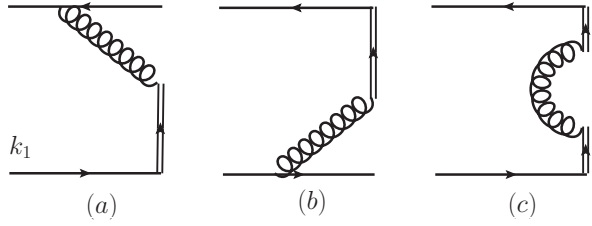}
\end{center}
\caption{The one-loop diagrams of the wave function. The double line represents the gauge link in Eq.(\ref{def}). }
\label{Feynman-dg1}
\end{figure}
\par
\par
We take the axial gauge to examine this. The gauge is fixed by $v\cdot G=0$ with the vector $v^\mu =(v^+,v^-,0,0)$.
$v^+$ and $v^-$ are arbitrary.
In the gauge the gluon propagator is given by:
\begin{equation}
\frac{-i}{q^2 +i\varepsilon} \left [ g^{\mu\nu} -\frac{v^\mu q^\nu +v^\nu q^\mu}{v\cdot q}
 + v^2 \frac{ q^\mu q^\nu }{(v\cdot q)^2} \right ].
\end{equation}
At one-loop, the wave function calculated with the off-shell quark pair is linearly related
to the gluon propagator because only one gluon is exchanged. Therefore, the possible extra
$\zeta$-dependence can only come from the second term in the above. The possible contributions
to this are represented by diagrams given in Fig.1 and Fig.2. We denote the one-loop contribution
from the second term to the wave function as $\phi^{(1)}_v$. The calculations are simple,
e.g., the contribution from Fig.1c:
\begin{equation}
\phi^{(1)}_v\biggr\vert_{1c} = -i g_s^2 \delta(x-x_0) \delta^2(\vec k_T-\vec k_{1\perp})
C_F \int\frac{d^4 q}{(2\pi)^4} \frac{u\cdot v {\rm Tr} \left (\gamma^- \gamma_5
\gamma^+\gamma_5 \gamma\cdot (k_1 -q) \gamma\cdot q \right ) }
 { (u\cdot q-i\varepsilon)(q^2 +i\varepsilon)((k_1-q)^2+i\varepsilon) (v\cdot q) },
\end{equation}
where we used for the external lines of the off-shell quark pair the spin projection
discussed before.  For the gluon propagator in Fig.1c we only used
a part in the second term in Eq.(8). Only this part deliveries the extra $\zeta$-dependence.
A caution should be taken for the denominator $1/v\cdot q$ in the above when one integrates
over $q^-$. One needs to supply a prescription for the denominator. This can give in the integration
an additional
pole than the pole from the first term. But the contribution from the additional pole does not depend
on $\zeta$ in the limit $\zeta\to\infty$. In principle there should be a transverse gauge link at $z^-=\infty$
in Eq.(7) to make the definition completely gauge invariant. In our case as long as we keep $u^+$ and $v^+$ nonzero,
the transverse gauge link will not introduce any contribution.
\par
It is straightforward to perform the loop integral in Eq.(9) and other loop integrals in Fig.1.
In the calculation
we will meet possible I.R. divergences. We introduce a small mass $\lambda$ for gluons
to regularize the I.R. divergences.
We have from
Fig.1:
\begin{eqnarray}
\zeta \frac{\partial}{\partial \zeta} \phi^{(1)}_v(x,k_T, \zeta, \mu)\biggr\vert_{1a}
&=& \zeta \frac{\partial}{\partial \zeta} \phi^{(1)}_v(x,k_T, \zeta, \mu)\biggr\vert_{1b} =
\phi^{(0)}(x,k_T, \zeta, \mu) \frac{\alpha_s C_F}{2\pi} \ln\frac{\lambda^2}{\mu^2},
\nonumber\\
\zeta \frac{\partial}{\partial \zeta} \phi^{(1)}_v(x,k_T, \zeta, \mu)\biggr\vert_{1c}
&=& -\phi^{(0)}(x,k_T, \zeta, \mu) \frac{\alpha_s C_F}{\pi} \left [ \ln\frac{\zeta^2 x^2}{k^2_T}
  +\frac{1}{2} \ln\frac{\zeta^2 x^2}{\mu^2} \right ]  ,
\nonumber\\
\zeta \frac{\partial}{\partial \zeta} \phi^{(1)}_v(x,k_T, \zeta, \mu)\biggr\vert_{1d}
&=&-\phi^{(0)}(x,k_T, \zeta, \mu) \frac{\alpha_s C_F}{\pi} \left [ \ln\frac{\zeta^2(1-x)^2}{k^2_T}
  +\frac{1}{2} \ln\frac{\zeta^2 (1-x)^2}{\mu^2} \right ],
\end{eqnarray}
For the contributions from Fig.2 it is easy to show that the sum does not depend on $\zeta$:
\begin{equation}
\zeta \frac{\partial}{\partial \zeta} \left [ \phi^{(1)}_v(x,k_T, \zeta, \mu)\biggr\vert_{2a}
+\phi^{(1)}_v(x,k_T, \zeta, \mu)\biggr\vert_{2b}+\phi^{(1)}_v(x,k_T, \zeta, \mu)\biggr\vert_{2c}\right ]
=0.
\end{equation}
Therefore we have in the gauge $v\cdot G=0$ for the wave function and the hard part $H$ at one-loop:
\begin{eqnarray}
\zeta \frac{\partial}{\partial \zeta} \phi(x,k_T, \zeta, \mu) &=&
  \left [ V \otimes \phi \right ] +\frac{\alpha_s C_F}{2\pi} \left [ 2 \ln\frac{\lambda^2}{k_T^2}
    - 3\ln\frac{\zeta^2 x^2}{k^2_T} -3\ln\frac{\zeta^2 (1-x)^2}{k^2_T} \right] \phi(x,k_T, \zeta, \mu),
\nonumber\\
\zeta \frac{\partial}{\partial \zeta} H(x,k_T, \zeta) &=&
-\left [ V \otimes  H \right ] - \frac{\alpha_s C_F}{2\pi} \left [ 2 \ln\frac{\lambda^2}{k_T^2}
    - 3\ln\frac{\zeta^2 x^2}{k^2_T} - 3\ln\frac{\zeta^2 (1-x)^2}{k^2_T} \right] H(x,k_T, \zeta),
\end{eqnarray}
where we use the notation $V \otimes \phi$ and $V \otimes  H $  to denote the contributions
from Feynman gauge, i.e., the contributions from the first term in Eq.(8).
The difference of signs in the two evolutions reflects the fact that the form factor
does not depend on $\zeta$.
From the above results, it is clear that the hard part $H$, hence the $k_T$-factorization, is
gauge-dependent, because it is different in different gauges. In the axial gauge,
the extra $\zeta$-dependence is I.R. divergent because of the contributions from Fig.1a and 1b.
Since the hard part receives the $\zeta$-dependence only from the wave function, the hard part in fact
contains a $\zeta$-dependent I.R. divergence.
This leads to the conclusion that the factorization is violated in this gauge.

\par\vskip10pt
{\bf 3.} In the general covariant gauge the hard part will receive a soft divergence called
light-cone divergence, as shown in \cite{FMW0}. In \cite{LiM} a method to eliminate this divergence
is suggested. But, this method is inconsistent as pointed out in \cite{FMW1}. Here, we explain
the inconsistence in detail.
\par
\begin{figure}[hbt]
\begin{center}
\includegraphics[width=5cm]{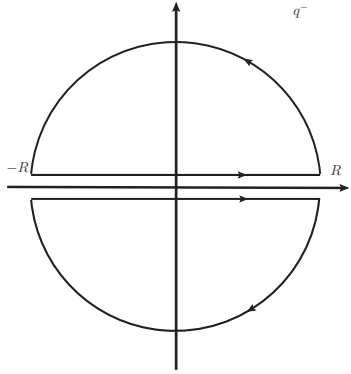}
\end{center}
\caption{The contours for the integration of $q^-$.  }
\label{Feynman-dg1}
\end{figure}
\par
The gauge-dependent contributions at one-loop are proportional to $\alpha$ in Eq.(4).
We denote these contributions as $\phi_\alpha$.
We take Fig.2b as an example. The gauge-dependent part is\cite{FMW0,LiM}:
\begin{equation}
\phi_\alpha(x,k_T) \biggr \vert_{2b}= \frac{ 8i \alpha \alpha_s}{\pi^2} \int_{-\infty}^{\infty} \frac{d q^-}{2\pi}
   \frac{ 2 (k_1^+-q^+) q^- -\vec k_{1\perp} \cdot  \vec q_\perp + q_\perp^2 }
      { [(k_1-q)^2 +i\varepsilon]
    ( q^2 + i\varepsilon )^2}
      \;.\label{2b}
\end{equation}
In the above $q$ is the momentum carried by the gluon in Fig.2b. The components
$q^+$ and $q_\perp^\mu$ are fixed as $q^+ = k_1^+ -x P^+$ and $\vec q_\perp = \vec k_{1\perp} -\vec k_T$
with $k_1^+ =x_0 P^+$.
The integral can simply be calculated by taking a closed contour in the upper-half- or in the lower-half complex
plan of $q^-$, as showing in Fig.3, where we take the contours consisting of a straight line along
the real axis and a semi circle with the radius $R$. The limit $R\to\infty$ should be taken
corresponding to that the integration over $q^-$ is from $-\infty$ to $\infty$.
From positions of $q^-$-poles of the integrand the integral only becomes nonzero in the region $0\leq q^+\leq x_0 P^+$,
because in this region, the pole from the $1/[(k_1-q)^2 +i\varepsilon]$ is in the upper-half
plan and the double pole from $1/( q^2 + i\varepsilon )^2$ is in the lower-half plan.
For the region $q^+ <0$ all poles are above the real axis, while all pole are below the real axis
for $q^+ >x_0 P^+$.
Taking any one of the two contours the integral can be performed. The result does not depend
on the choice of contours.
\par
It is found in \cite{FMW0} that the result is singular. To clearly
see this, we take a test function $t(x,k_\perp)$ to calculate the
convolution $t\otimes \phi^{(1)}$. By taking $t(x,k_\perp)=
H^{(0)}(x,k_\perp,y_0,k_{2\perp})$ one obtains the convolution in
Eq.(6). The singularity can be regularized by a small gluon mass
$\lambda_L$ or with dimensional regularization as showing in
\cite{FMWB}. We first discuss the case with the dimensional
regularization, in which the transverse space is $2-\epsilon_L$
dimensional with $\epsilon_L \to 0$. We have then  the singular
contribution of $\phi_\alpha\vert_{2b}$ and the convolution:
\begin{eqnarray}
\phi_\alpha\biggr \vert_{2b} &=&  -\frac {4\alpha \alpha_s }{\pi^2}
   \frac{ k^2_{1\perp} \theta(q^+) \theta (k_1^+ -q^+) }
   {   k_1^+ ( q^2_\perp + q^+ k^2_{1\perp} /k_1^+ )^2 } +{\rm finite\ terms},
\nonumber\\
 t\otimes \phi_\alpha\biggr\vert_{2b} & = &  \frac{ 8 \alpha
\alpha_s }{\pi P^+}  t(x_0,k_{1\perp} )
  \left ( \frac{1}{\epsilon_L} \right ) +
   ({\rm finite\ terms }).
\end{eqnarray}
From Eq.(13) it is clear that the singularity comes from the momentum region where
$q$ has the scaling patten $q^\mu \sim (\delta_0^2, 1, \delta_0,\delta_0)$ with $\delta_0\ll 1$.
The singularity comes from the first term in Eq.(13).
We call this as light-cone singularity. Adding all contributions at one-loop
the singularity is not canceled. It results in that the hard part $H$ at one-loop
is divergent in the general covariant gauge. It should be noted that the form factor
does not contain such singularities\cite{FMW0,FMWV1}.
\par
To deal the singularity a method is suggested in \cite{LiM}. In the method one keeps
$R$ large but finite in the calculation of the wave function and the limit $R\to\infty$
is only taken after that the integrals in the convolution are performed.
Then one obtains for the wave function not only the contribution from the pole but
also the contribution from the semi-circle. As suggested in \cite{LiM} for the contribution from Fig.2b,
one takes the contour
for $0 < q^+$ in the upper-half plan and the contour for $q^+ <0$
in the lower half plan, one then has:
\begin{eqnarray}
\phi_\alpha\biggr \vert_{2b}   &=& -\frac {4\alpha \alpha_s }{\pi^2}
   \frac{ k^2_{1\perp} \theta(q^+) \theta (k_1^+ -q^+) }
   {   k_1^+(  q^2_\perp + q^+ k^2_{1\perp}/k_1^+ )^2 }
    -\frac{4\alpha \alpha_s}{\pi^3}
\left[  \theta(q^+) \int_\pi^0 d\theta
+ \theta(-q^+) \int_{-\pi}^0 d\theta   \right ] {\mathcal F}_R (\theta, x,k_\perp) +\cdots,
\nonumber\\
  && {\mathcal F}_R (\theta, x,k_\perp) =
\frac{2(k_1^+-q^+)(Re^{i\theta})^2}{(2q^+Re^{i\theta}-q_\perp^2)^2
[2(q^+-k_1^+)Re^{i\theta}-|{\vec k}_{1\perp}-{\vec q}_\perp|^2]},
\end{eqnarray}
where ${\mathcal F}_R $ is the first term in Eq.(12) with $q^-= Re^{i\theta}$.
There are finite terms denoted as $\cdots$.
 The terms
in the second line are from semi circles. If we take the limit $R\to\infty$
in the above as required in Eq.(13), we obtain the result in Eq.(14). This also results in that
$\phi_{\alpha}\vert_{2b}$ is nonzero only in the region $0< q^+< k_1^+$ or $ 0 < x<x_0$. As suggested in \cite{LiM}
one should keep $R$ finite here and calculate the convolution first. The limit is taken
after the integrations in the convolution. In this case one has:
\begin{eqnarray}
t\otimes \phi_\alpha\biggr\vert_{2b} &=& \frac{ 8 \alpha \alpha_s }{\pi P^+}  t(x_0,k_{1\perp} )
 \left ( \frac{1}{\epsilon_L} \right ) +\cdots
\nonumber\\
  &&  - \frac{4 \alpha \alpha_s}{\pi^3 P^+} \lim_{R\to\infty}\int d^2q_\perp
\left[\int_\pi^0 d\theta\int_0^{k_1^+} dq^+
+\int_{-\pi}^0 d\theta\int_{-\infty}^0 dq^+\right] t(x,k_\perp) {\mathcal F}_R (\theta,x,k_\perp),
\label{con}
\end{eqnarray}
Working out the singular part of the integrals in the second line and taking the limit $R\to \infty$  one has:
\begin{eqnarray}
t\otimes \phi_\alpha\biggr\vert_{2b} = \frac{ 8 \alpha \alpha_s }{\pi P^+}  t(x_0,k_{1\perp} )
 \left ( \frac{1}{\epsilon_L} \right )
  + \frac{4\alpha \alpha_s}{P^+ \pi^2}
\left[\int_\pi^0 d\theta
 - \int_{-\pi}^0 d\theta\right]  t(x_0,k_{1\perp} )  \left ( \frac{1}{\epsilon_L} \right )  + ( {\rm finite\ terms }) .
\end{eqnarray}
From the above the convolution calculated with the method from \cite{LiM} is free from the singularity.
Two observations can be made at the first look from the above result.
Because the limit $R\to\infty$ is taken after the integrals in the convolution, it implies
that one introduces a cut-off for $q^-$ with $\vert q^- \vert < R$. It is unclear how to implement
the cut-off in the definition in Eq.(\ref{def}). The finite $R$ results in
that the contribution is not zero with $q^+ <0$ corresponding to $x>x_0$.
Another observation
is that the result depends on contours. Different choices of contours give different results.
\par
For the region with $0 < q^+ $ or $q^+ <0$ one always has two possibilities
of contours, respectively. In the above the contour for $0 < q^+$
is in the upper-half plan. If we take the contour in the lower-half plan with the finite $R$
for $0 < q^+$ and $ 0 > q^+$,
we then have:
\begin{eqnarray}
\phi_\alpha\biggr \vert_{2b} = -\frac {4\alpha \alpha_s }{\pi^2}
   \frac{ k^2_{1\perp} \theta(q^+-q_0^+) \theta (k_1^+ -q^+) }
   {   k_1^+ ( q^2_\perp + q^+ k^2_{1\perp}/k_1^+ )^2 }
   -\frac{4\alpha \alpha_s}{\pi^3}
\left[  \theta(q^+) \int_{-\pi}^0 d\theta
+ \theta(-q^+) \int_{-\pi}^0 d\theta   \right ] {\mathcal F}_R (\theta, x,k_\perp) +\cdots,
\end{eqnarray}
where $q_0^+ = q^2_\perp/(2R)$. Here the contribution from the double pole starts at $q^+ =q_0^+$
instead of $q^+ =0$, because the double pole with small enough $q^+$ can be in the outside
of the contour with $R$. In the above the first $\theta$-integral is
from $\theta=-\pi$ to $\theta =0$ because the contour here is in the lower-half plan.
Comparing with Eq.(15), one realizes that $\phi_\alpha\vert_{2b}$ is different with different
contours. This in turn gives different results of the convolution.

\par
Analyzing all possibilities one can have three different results for the following
choices of contours in the two regions of $q^+$: (I). All contours are in the upper-half plan or in the lower-half
plan.
(II). The contour for $0< q^+ $ is in the upper-half plane and
the contour for $q^+ <0$ is in the lower-half plan.
(III). The contour for $0< q^+ $ is in the lower-half plane and
the contour for $q^+ <0$ is in the upper-half plan.
The choice (II) corresponding to Eq.(15,16).
The different results for these 3 choices are summarized as:
\begin{eqnarray}
t\otimes \phi_\alpha\biggr\vert_{2b} &=&\frac{ 8 \alpha \alpha_s }{\pi P^+}  t(x_0,k_{1\perp} )
  \left ( \frac{1}{\epsilon_L} \right ) +( {\rm finite\ terms }),  \ \ \  {\rm for\ (I)},
\nonumber\\
t\otimes \phi_\alpha\biggr\vert_{2b} &=&\frac{ 8 \alpha \alpha_s }{\pi P^+}  t(x_0,k_{1\perp} )
\left ( \frac{1}{\epsilon_L} \right )  -\frac{ 8 \alpha \alpha_s }{\pi P^+}  t(x_0,k_{1\perp} )
\left ( \frac{1}{\epsilon_L} \right )
   +( {\rm finite\ terms }),  \ \ \  {\rm for\ (II)},
\nonumber\\
t\otimes \phi_\alpha\biggr\vert_{2b} &=&
\frac{ 8 \alpha \alpha_s }{\pi P^+}  t(x_0,k_{1\perp} )
\left ( \frac{1}{\epsilon_L} \right )  + \frac{ 8 \alpha \alpha_s }{\pi P^+}  t(x_0,k_{1\perp} )
\left ( \frac{1}{\epsilon_L} \right )
+( {\rm finite\ terms }),  \ \ \  {\rm for\ (III)}.
\end{eqnarray}
This is the contour dependence pointed out in \cite{FMW1}. The cancelation of the singularity depends
on the choice of contours with the method in \cite{LiM}.
\par
One may use a small gluon mass $\lambda$ to regularize the singularity. The mass in the general covariant
gauge is introduced as in \cite{IZ}:
\begin{equation}
\frac{-i g^{\mu\nu}}{q^2 -\lambda^2+ i\varepsilon}
 +i \alpha \frac{ q^\mu q^\nu}{(q^2-(1- \alpha) \lambda^2+i\varepsilon )(q^2 -\lambda^2+ i\varepsilon) }.
\end{equation}
Comparing with Eq.(4), the double pole in the $q^-$-plan splits into two single poles.
It can be shown with the method in \cite{LiM} that the singularity is canceled by the contributions
from semi circles. It is interesting to note that the cancelation is contour-independent,
because an additional contribution from the cases where one of the two single poles
can be in the outside of a closed contour with the finite $R$. However, this only works at one-loop,
because one has at one-loop only diagrams similar to QED. Beyond one-loop level, one can not
take a finite gluon mass here for QCD.
\par
A natural question with the suggest
method in \cite{LiM} is then which answer is correct? When one takes the dimensional regularization,
the result is contour-dependent. When one uses a finite gluon mass for the regularization,
it is definitely not correct beyond one-loop level.
From the above analysis different results
about the singularity are due to keeping $R$  finite in Eq.(15) instead of $R\to \infty$.
In fact, none of them is correct
in the sense that $R$ can not be kept finite in the wave function.
A finite $R$ is in conflict with translational covariance.
Below we discuss the conflict in detail.
\par
We start with the well-known fact that the wave function becomes zero with $x> 1$.
This is derived by sandwiching a complete set of states into Eq.(\ref{def}) and using the
translational covariance. It should be noted that $x_0$ in Eq.(\ref{mom}) is arbitrary in the region
$ 0< x_0 <1$. For nonzero $k_{1\perp}$ we can take $x_0 =1$ for the off-shell quark pair $q(k_1) \bar q(P-k_1)$
as an extreme case without any problem, because
the quark and the antiquark with $x_0=1$ still have nonzero momenta. Then, with a large but still finite $R$,
the wave function is not zero with $x >1$. This can also be seen from the contributions in Eq.(15).
Therefore, a finite $R$ is in conflict
with translational covariance in the case with $x_0=1$. In general, as showing in \cite{FMW1},
the contributions to the wave function from a class of diagrams including Fig.2b, where gluons are only exchanged
between the quark field $q(z)$ and $L_u^\dagger (\infty,0)$ in Eq.(\ref{def}) or are exchanged
to form self-energy corrections to $q(z)$ and $L_u^\dagger (\infty,0)$, are proportional to
\begin{eqnarray}
  \int \frac{ d z^- }{2\pi}
  \frac {d^2 z_\perp}{(2\pi )^2}  e^{ix P^+ z^- - i \vec z_\perp\cdot \vec k_T}
\langle 0 \vert \left (  L_u^\dagger (\infty, 0) - 1 \right )
  \gamma^+ \gamma_5  q(z) \vert q(k_1) \rangle\vert_{z^+=0}
\label{deff}
\end{eqnarray}
besides some trivial factors.
The leading order result of the above expression is given by Fig.2b by deleting the antiquark line.
Sandwiching a complete set of states and using the
translational covariance, it is easy to show  that the above quantity
must be zero with $x> x_0$.
Here $x_0$ is arbitrary in the region $0< x_0 <1$.
Therefore, the contribution from Fig.2b must be zero with  $x> x_0$.
A nonzero contribution from Fig.2b with $x> x_0$ is unphysical.
Taking $\vert q^- \vert <R$ instead of $\vert q^- \vert < \infty$ in Eq.(13,15), it results in
that the contribution is not zero with $x>x_0$.
One also notes from Eq.(15,18) that with the finite $R$ the wave function is not zero
for $q^+ > k_1^+$ or $x <0$. It implies that the quark entering hard scattering is with
a negative energy.
This is inconsistent not only with the translational covariance but also with physical picture.
\par
From the above discussion, it is clear that
the method of \cite{LiM}
by keeping $R$ finite in the wave function is essentially to include
the unphysical contributions. Convoluting the wave function containing these unphysical
contributions with a test function, the discussed light-cone singularities
may be canceled. The cancelation depends on contours and also on how the
singularities are regularized. As shown in the above,
the existence of these unphysical contributions is in conflict with general principles,
i.e., with the translational
covariance.
Because of this
and the contour dependence the method is not consistent.
We emphasize here that the key problem of the method is the inconsistence
with the translational covariance for the defined
wave function in Eq.(7). This in fact can be easily seen by noting
the following fact: The finite cut-off $R$ for $q^-$ implies that
one takes the corresponding space-time coordinate $x^+$ as one-dimensional lattice with the
lattice spacing $2\pi/R$. The lattice certainly has no symmetry of translational covariance.
The conclusion here is, as made in \cite{FMW0,FMW1},
that the hard part will contain the divergent part in the general covariant gauge.  Hence
the $k_T$-factorization is gauge-dependent and violated in this gauge.

\par
To summarize: With the results presented in this note, one can conclude
that the $k_T$-factorization is gauge-dependent and violated in the gauges studied here.
Since some sources of the gauge dependence studied here are from wave functions and
absent in scattering amplitudes of off-shell partons, the $k_T$-factorization
is gauge-variant for any exclusive process. Our results can be generalized to the case
of $B$-meson decays. Therefore, the $k_T$-factorization for exclusive
$B$-meson decays is also gauge-dependent.

\par\vskip20pt
\noindent
{\bf Note added:}
\par
After we completed the present work, an analysis of three-parton contribution to the
$\pi$-form factor in the $k_T$-factorization appeared in \cite{ChLi}.
The contribution is at twist-three, where the spin projection
for the initial- and final off-shell quark pair are
$\sigma^{- \alpha} \gamma_5$ and $\sigma^{+\beta}\gamma_5$, respectively.
In \cite{ChLi} the $k_T$-factorization has also been studied in the general covariant gauge.
Because
the different spin projections, the results obtained in \cite{ChLi} has no implication
for our result in Eq.(5) and our discussion there. But the results with the general covariant
gauge in \cite{ChLi} clearly show that the factorization is gauge-variant.
\par
From the results in \cite{ChLi} the gauge dependent part in the three-parton contribution
related to the initial $\pi$ can be written in the form of the following
convolution:
\begin{eqnarray}
\alpha  H_{\mu\nu}  \otimes \langle 0 \vert \bar q(y) \gamma_5
\gamma^+ \gamma^\mu D^{\nu}  (0) q(0)\vert \pi (P)\rangle,
\end{eqnarray}
where $\alpha$ is the gauge parameter in Eq.(4) and
$H^{\mu\nu}$ is perturbative coefficient function.
The field $\bar q$ is with $y^\mu
=(0,y^-,\vec y_\perp)$. It should be noted $\vec y_\perp\neq 0$
because the quark carries a nonzero transverse momentum into
hard scattering in the $k_T$-factorization. For $\vec y_\perp =0$
the above matrix-element is of twist-3 in collinear factorization.
It has been argued in \cite{ChLi} that the above
expression is zero because of the equation of motion. But it is not
true. There are two possibilities to have the above quantity to be
zero with equation of motion. If one has $H^{\mu\nu} \propto g^{\mu\nu}$, it is clear that
the quantity is zero.
In general, the perturbative coefficient functions related
to the derivative and to the gluon field in the covariant derivative
$D^{\nu}(x) =\partial^{\nu} + ig_s G^{\nu}(x)$ can be different.
In \cite{ChLi} one has shown that the contribution related to the $\partial^+$
can be neglected and the attachment of a $+$-component of gluon to the
hard scattering diagrams gives no contribution. From these results
one has $H^{+ - } =0$.
Therefore, $H^{\mu\nu}$ is not
proportional to $g^{\mu\nu}$. Another possibility is that the matrix
element is proportional to $g^{\mu\nu}$. But again this is not
correct. To show this, one can make a Fourier transformation of
$\langle 0 \vert \bar q(y) \gamma_5 \gamma^\rho \gamma^\mu  D^{\nu}
(0) q(0)\vert \pi (P)\rangle$ and study its decomposition.
The decomposition is complicated. The first few terms are:
\begin{eqnarray}
  && P^+ \int\frac{dy^- d^2 y_\perp}{(2\pi)^3}
  e^{ -i x P^+ y^- + i \vec k_\perp\cdot \vec y_\perp}
\langle 0 \vert \bar q(y) \gamma_5 \gamma^\rho \gamma^\mu  D^{\nu}
(0) q(0)\vert \pi (P)\rangle
\nonumber\\
&& =  f_0 (x, k_\perp) \left ( P^\rho g^{\mu\nu} -g^{\rho\nu} P^\mu \right )
  + f_1 (x,k_\perp) g^{\rho\mu}P^\nu + f_2(x,k_\perp)  \left ( g^{\rho\nu}k_\perp^\mu
     -g^{\mu\nu}k_\perp^\rho \right ) + f_3 (x,k_\perp) g^{\rho\mu} k_\perp^\nu
\nonumber\\
  && \ \ \ +  f_4(x,k_\perp ) \left ( P^\rho k_\perp^{\mu} k_\perp^{\nu} -P^\mu k_\perp^{\rho} k_\perp^{\nu} \right )
   + \cdots,
\end{eqnarray}
where $\cdots$ denote the structures which are zero when $\rho=+$ and those may depend
on $n^\mu =(0,1,0,0)$. In
general the functions $f_{0,1,2,\cdots}$ are not zero. From the reason of dimension
they may be at the same level of importance. This can also be seen the constraint of $f$'s
obtained by contracting
$g_{\mu\nu}$ from both sides.
Taking $\rho =+$, one clearly sees that the
matrix element is not proportional to $g^{\mu\nu}$. Therefore, the
above quantity is not zero. This shows that the three-parton
contribution in the $k_T$-factorization for the $\pi$-form factor in \cite{ChLi} is
gauge-dependent.

\vskip 5mm
\par\noindent
{\bf Acknowledgments}
\par
This work is supported by National Nature
Science Foundation of P.R. China(No. 10805028, 10975169, 11021092).
\par\vskip30pt



\begin{thebibliography}{99}

\bibitem{NLi} S. Nandi and H.-n. Li, Phys. Rev. D76 (2007) 034008, e-Print: arXiv:0704.3790 [hep-ph].

\bibitem{FMW0} F. Feng, J.P. Ma and Q. Wang, Phys. Lett. B674 (2009) 176,
e-Print: arXiv:0807.0296 [hep-ph].

\bibitem{LiM} H.-n. Li and S. Mishima, Phys. Lett. B674 (2009) 182, e-Print: arXiv:0808.1526 [hep-ph].

\bibitem{FMW1} F. Feng, J.P. Ma and Q. Wang, Phys. Lett. B677 (2009) 121,
e-Print: arXiv:0808.4017 [hep-ph].

\bibitem{LiS} H.-n. Li, Y.L. Shen, Y.M. Wang and H. Zou, e-Print: arXiv:1012.4098 [hep-ph].

\bibitem{WZT} Z.T. Wei, private communication.

\bibitem{TMD1} J.P. Ma and Q. Wang, Phys. Lett. B642 (2006) 232, hep-ph/0605075.

\bibitem{FMWB} F. Feng, J.P. Ma and Q. Wang, e-Print: arXiv:0901.2965 [hep-ph].

\bibitem{FMWV1} F. Feng, J.P. Ma and Q. Wang, e-Print: arXiv:0808.4017v1 [hep-ph].

\bibitem{IZ} C. Itzykson and J. Zuber, {\it Quantum Field Theory}, New York, McGraw-Hill (1980).

\bibitem{ChLi}
Y.-C. Chen and H.-n. Li, e-Print: arXiv:1104.5398 [hep-ph].





\end{thebibliography}
\end{document}